\documentclass[epj,nopacs]{svjour}
\usepackage{graphics}
\usepackage{graphicx}
\begin{document}
\title{New Results on $\tau$ Lepton}

\author{K.K. Gan
}                     
%
%
\institute{Department of Physics,
        The Ohio State University,
        Columbus, OH 43210,
        U.S.A.}%
\date{Received: date / Revised version: date}
%
\abstract{
This is a review of new results on the $\tau$ lepton.
The lifetime has been measured with improved precision,
allowing the test of lepton universality to 0.2\% level.
There are several new measurements of the branching ratios for the
hadronic decays, including much improved measurements for the
Cabibbo-suppressed decays.
All results are consistent with the Standard Model expectations.
\PACS{
      {PACS-key}{discribing text of that key}   \and
      {PACS-key}{discribing text of that key}
     } 
} 
\maketitle
\section{Introduction}
\label{sec:intro}
The $\tau$ lepton provides a unique laboratory to test the Standard Model.
Its large mass allows the lepton to decay into both lighter leptons and hadrons,
providing many venues to challenge the Standard Model.
This includes the test of lepton universality,
Conserved-Vector-Current (CVC) hypothesis~\cite{Feynman}, and isospin symmetry.
In this paper, I will first review the new measurements of the lifetime
and then the branching ratios for hadronic decays.

\section{Lifetime}
\label{sec:Lifetime}
In the Standard Model, the coupling of the $\tau$ lepton to the $W$
boson is the same as those for the lighter leptons, $e$ and $\mu$.
The universality of the couplings can be tested by comparing the
measurement of the $\tau$ lifetime ($\tau_\tau$) and leptonic
branching fractions~\cite{Tsai}:
\begin{equation}
\tau_\tau = \tau_\mu\left (\frac{g_\mu}{g_\tau}\right )^2
                    \left (\frac{m_\mu}{m_\tau}\right )^5
            B(\tau^- \to e^-{\bar \nu_e}\nu_\tau)
\end{equation}
\begin{equation}
\tau_\tau = \tau_\mu\left (\frac{g_e}{g_\tau}\right )^2
                    \left (\frac{m_\mu}{m_\tau}\right )^5
            B(\tau^- \to \mu^-{\bar \nu_\mu}\nu_\tau)
           f\left (\frac{m^2_\mu}{m^2_\tau}\right )
\end{equation}
\noindent
where the phase space factor is given by
\begin{equation}
f(x) = 1 - 8x + 8x^3 - x^4 -12x \ln x = 0.9726,
\end{equation}

\noindent
where $\tau_\mu$ is the $\mu$ lifetime, $m_\mu$ and $m_\tau$ is
the mass of the $\mu$ and $\tau$, and $g_e$, $g_\mu$ and $g_\tau$ is
the coupling of the $W$ boson to $e$, $\mu$, and $\tau$.
Lepton universality corresponds to $g_e = g_\mu = g_\tau$.

Both DELPHI and BaBar have reported preliminary measurements of the $\tau$ lifetime.
The DELPHI experiment~\cite{DELPHI1} used three techniques to measure the lifetime:

$\bullet$ 3-prong decay length

$\bullet$ 1-prong vs 1-prong impact parameters:

\hskip 0.2 in $\bullet$ impact parameter difference: $d_1 - d_2$

\hskip 0.2 in $\bullet$ miss distance: $d_1 + d_2$

\noindent The measurements of lifetime using the two impact parameter
techniques have 36\% correlation and this has been taken into account
by the experiment to extract a new measurement of the lifetime.

BaBar has reported a preliminary measurement of
the lifetime using the 3-prong decay length method.
The experiment performed a blind analysis which is highly
desirable for this kind of high precision measurement.

\begin{figure}
\centering
\includegraphics[width=6.8cm]{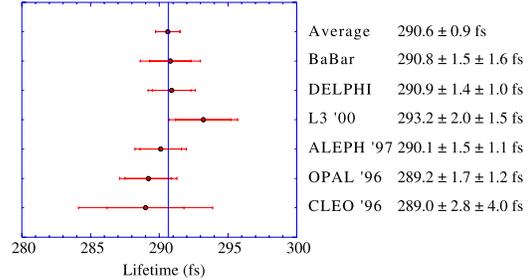}
\caption{Measurements of the $\tau$ lifetime.}
\label{fig:lifetime}
\end{figure}

The current status of the lifetime measurements is shown
in Fig.~\ref{fig:lifetime}.
It is reassuring to see that all measurements are consistent
with the result from a blind analysis.
There is a good prospect for higher precision measurements by
BaBar and Belle given their large data samples.

\begin{figure}
\centering
\includegraphics[width=4.1cm]{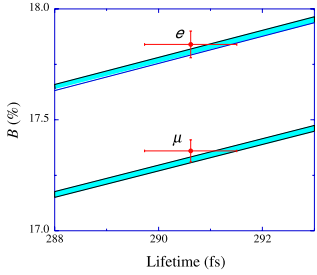}
\caption{Measurements of the leptonic branching ratios vs $\tau$ lifetime.
The lines show the predictions based on lepton universality with the
width given by the uncertainty in the $\tau$ mass.}
\label{fig:B_lepton}
\end{figure}

The average measured lifetime is plotted against the average
measured leptonic branching ratios~\cite{PDG} in Fig.~\ref{fig:B_lepton}.
Also shown are the Standard Model predictions assuming lepton universality.
The measurements are consistent with lepton universality.
Alternatively we can use the measurements to extract the following
ratios of couplings:
\begin{eqnarray*}
\frac{g_\mu}{g_\tau} = 0.9990 \pm 0.0023\\
\frac{g_e}{g_\tau}   = 0.9988 \pm 0.0021
\end{eqnarray*}
\noindent
Both ratios are consistent with the Standard Model expectation to a
precision of 0.2\%, a remarkable achievement.

\section{Hadronic Decays}
\label{sec:Hadronic}
The $\tau$ lepton can decay into many hadronic states due to its large mass.
The large number of hadronic channels allows the study of many aspects
of the Standard Model, including the test of lepton universality,
CVC, and isospin symmetry.
In this section, I will review the new results from DELPHI~\cite{DELPHI2},
L3~\cite{L3}, and CLEO~III~\cite{CLEO}.
The first two experiments measure the inclusive branching ratios while
CLEO~III measures the exclusive branching ratios with $\pi$/$K$ identification.

\subsection{Branching ratio of $\tau^- \to h^-\nu_\tau$}
\label{sec:h}
\begin{figure}
\centering
\includegraphics[width=6.8cm]{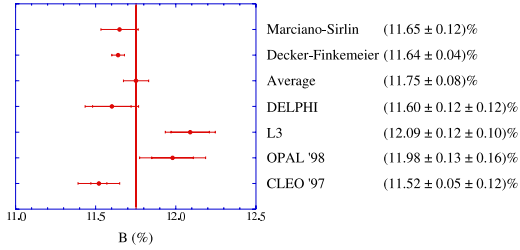}
\caption{Branching ratio for $\tau^- \to h^-\nu_\tau$.}
\label{fig:B_h}
\end{figure}

The decay $\tau^- \to \pi^-\nu_\tau$ involves the coupling of the
weak current to the pion.
The branching ratio for $\tau^- \to \pi^-\nu_\tau$ can be
calculated~\cite{Tsai} from that of $\pi^- \to \mu^-\bar \nu_\mu$,
\begin{eqnarray}
\frac{B(\tau^- \to \pi^-\nu_\tau)}{B(\pi^- \to \mu^-\bar \nu_\mu)} 
=  \left[\frac{g_\tau}{g_{\mu}}\right]^2
\frac{\tau_\tau}{\tau_\pi}
\frac{1}{2} \frac{m^3_\tau}{m_\pi m^2_\mu}
\frac{(1- m^2_\pi/m^2_\tau)^2}{(1- m^2_\mu/m^2_\pi)^2}
\label{eq:B_pi}
\end{eqnarray}
where $\tau_\pi$ and $m_\pi$ is the $\pi$ lifetime and mass, respectively.
A measurement of the branching ratio provides a test of $\mu-\tau$ universality.
Most of the experiments performed only an inclusive
measurement, $B(\tau^- \to h^-\nu_\tau)$ with $h = \pi$ or $K$.
The new results~\cite{DELPHI2,L3} are shown in Fig.~\ref{fig:B_h}.
There is considerable spread among the measurements, with
$\chi^2$ of 9.9 for 3 DOF (CL = 2\%).
The average branching ratio is consistent with the
expectation~\cite{Marciano_hadron} from the
Standard Model, including radiative corrections.

\subsection{Branching ratio of $\tau^- \to h^-\pi^0\nu_\tau$}
\label{sec:hpi0}
\begin{figure}
\centering
\includegraphics[width=7.8cm]{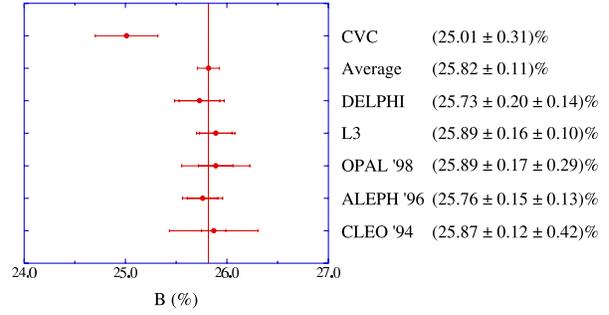}
\caption{Branching ratio for $\tau^- \to h^-\pi^0\nu_\tau$.}
\label{fig:B_hpi0}
\end{figure}

The decay $\tau^- \to \pi^-\pi^0\nu_\tau$ involves the coupling of the
weak vector current to the $\rho$ meson.
The branching ratio can be calculated~\cite{Tsai} by using the
CVC hypothesis to relate the coupling strength of the $\rho$
to the weak charged vector current and the electromagnetic
neutral vector current.
This isospin invariance of the quark current couplings to
the gauge bosons is a feature of the Standard model.
A measurement of the branching ratio for $\tau^- \to \pi^-\pi^0\nu_\tau$
therefore provides a test of the Standard Model.
Most of the experiments performed only an inclusive
measurement, $B(\tau^- \to h^-\pi^0\nu_\tau)$.
The new results~\cite{DELPHI2,L3} are shown in Fig.~\ref{fig:B_hpi0}.
Also shown is the theoretical expectation which is
a sum of the CVC predictions~\cite{Davier1} and the measured
branching ratio~\cite{PDG} for $\tau^- \to K^-\pi^0\nu_\tau$.
The CVC prediction is based on a reanalysis of the measured
cross section for $e^+e^- \to \pi^+\pi^-$, with a more proper
treatment of the radiative corrections.
All the $\tau$ measurements are consistent with each other but the
average is somewhat higher than the CVC expectation.

\subsection{Branching ratio of $\tau^- \to h^-2\pi^0\nu_\tau$}
\label{sec:h2pi0}
\begin{figure}
\centering
\includegraphics[width=7.8cm]{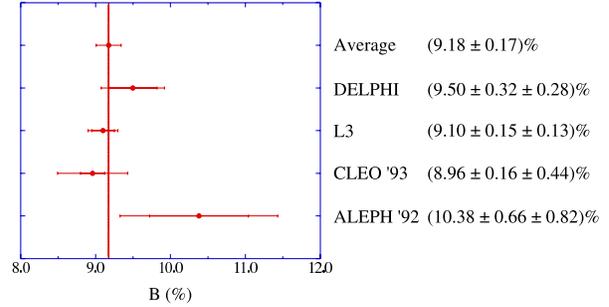}
\caption{Branching ratio for $\tau^- \to h^-2\pi^0\nu_\tau$.}
\label{fig:B_h2pi0}
\end{figure}

The current state of $B(\tau^- \to h^-2\pi^0\nu_\tau)$ is
summarized in Fig.~\ref{fig:B_h2pi0}, including the two new
measurements~\cite{DELPHI2,L3}.
All the measurements are consistent with each other.
Correcting for the small contribution of $\tau^- \to K^-2\pi^0\nu_\tau$
yields the exclusive branching ratio,
$B(\tau^- \to \pi^-2\pi^0\nu_\tau) = (9.12 \pm 0.17)\%$.

\subsection{Branching ratio of $\tau^- \to h^-h^+h^-\nu_\tau$}
\label{sec:3h}
\begin{figure}
\centering
\includegraphics[width=6.4cm]{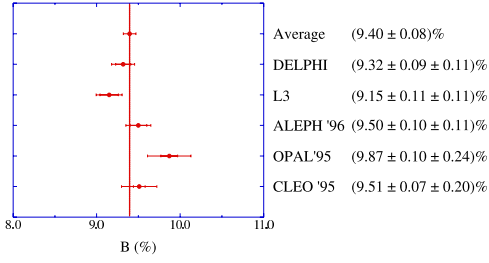}
\caption{Branching ratio for $\tau^- \to h^-h^+h^-\nu_\tau$.}
\label{fig:B_3h}
\end{figure}

The new results by DELPHI~\cite{DELPHI2} and L3~\cite{L3} on
$B(\tau^- \to h^-h^+h^-\nu_\tau)$ together with the results from
other experiments are summarized in Fig.~\ref{fig:B_3h}.
All the measurements are consistent with each other.

\subsection{Branching ratio of $\tau^- \to \pi^-\pi^+\pi^-\nu_\tau$}
\label{sec:3pi}
The CLEO~III experiment uses their ring imaging Cheren-kov detector (RICH)
to separate $\pi$/$K$ and obtains the first measurement of the exclusive
branching ratio~\cite{CLEO}:
$B(\tau^- \to \pi^-\pi^+\pi^-\nu_\tau) = (9.13 \pm 0.05 \pm 0.46)\%$.
Comparing this result with that in Sec.~\ref{sec:h2pi0},
it is evident that they  are consistent with isospin symmetry:
$B(\tau^- \to \pi^-2\pi^0\nu_\tau) \le
B(\tau^- \to \pi^-\pi^+\pi^-\nu_\tau)$.

\subsection{Branching ratio of $\tau^- \to K^-\pi^+\pi^-\nu_\tau$}
\label{sec:K2pi}
\begin{figure}
\centering
\includegraphics[width=6.4cm]{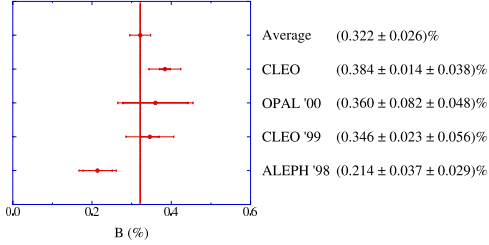}
\caption{Branching ratio for $\tau^- \to K^-\pi^+\pi^-\nu_\tau$.}
\label{fig:B_K2pi}
\end{figure}

The CLEO~III experiment has also measured the branching ratio for
$\tau^- \to K^-\pi^+\pi^-\nu_\tau$~\cite{CLEO}.
This result together with those of other experiments are summarized
in Fig.~\ref{fig:B_K2pi}.
The measurements are consistent with each other.

\subsection{Branching ratio of $\tau^- \to K^-K^+\pi^-\nu_\tau$}
\label{sec:KKpi}
\begin{figure}
\centering
\includegraphics[width=6.2cm]{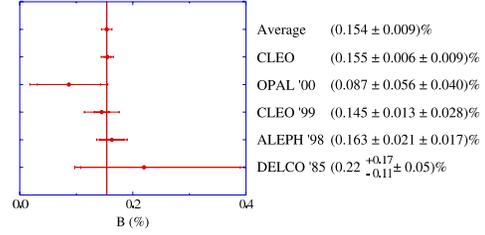}
\caption{Branching ratio for $\tau^- \to K^-K^+\pi^-\nu_\tau$.}
\label{fig:B_KKpi}
\end{figure}

CLEO~III has also measured the branching ratio for
$\tau^- \to K^-K^+\pi^-\nu_\tau$~\cite{CLEO}.
This result is consistent with other experiments but is significantly
more precise as shown in Fig.~\ref{fig:B_KKpi}.
The experiment also set an upper limit on the three-koan decay,
$B(\tau^- \to K^-K^+K^-\nu_\tau) < 0.0037\%$ at the 90\%CL.

The sum of the three exclusive 3-prong
branching ratios, $\approx (9.6 \pm 0.5)\%$, is consistent with
the inclusive result shown in Sect.~\ref{sec:3h},
indicating internal consistency of the measurements.

\subsection{Branching ratio of $\tau^- \to h^-h^+h^-\pi^0\nu_\tau$}
\label{sec:3hpi0}
\begin{figure}
\centering
\includegraphics[width=6.2cm]{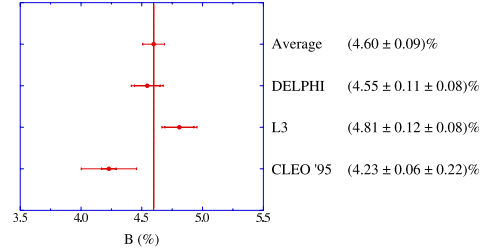}
\caption{Branching ratio for $\tau^- \to h^-h^+h^-\pi^0\nu_\tau$.}
\label{fig:B_3hpi0}
\end{figure}

The new results by DELPHI~\cite{DELPHI2} and L3~\cite{L3} on
$B(\tau^- \to h^-h^+h^-\pi^0\nu_\tau)$ together with the result by
CLEO are summarized in Fig.~\ref{fig:B_3hpi0}.
All the measurements are consistent with each other.
If we neglect the small contribution from decays with kaons,
the weighted average is significantly ($\approx 4\sigma$)
above the CVC prediction~\cite{Davier2}, $(3.63 \pm 0.21)\%$,
based on the measured cross section for $e^+e^- \to 4\pi$.
More precise measurements of the branching ratio and cross section
are needed to resolve the discrepancy.

\section{Conclusion}
\label{sec:Conclusion}
We have reached a new level of sensitivity
in $\tau$ physics.
Lepton universality is now tested to the level of 0.2\%.
We have begun a new era of $\tau$ physics with kaons and can
expect new results from CLEO~III, BaBar, and Belle.
There is no hint of physics beyond the Standard Model.

\vspace{0.1cm}
\noindent
{\bf Acknowledgements}\\
This work was supported in part by the U.S.~Department of Energy
under contract No. DE-FG-02-91ER-40690.
%
%

\end{document}